\documentclass[aps,twocolumn]{revtex4}
\usepackage{graphicx}
\begin{document}

\title{Charged Particles on Surfaces: Coexistence of Dilute Phases and Periodic Structures on Membranes}
\author{Sharon M. Loverde}\affiliation{Department of Materials Science and
Engineering, Northwestern University, Evanston, Illinois
60208-3108}
\author{Francisco J. Solis}\affiliation{Integrated Natural Sciences, Arizona State University West,
Integrated Natural Sciences, Phoenix, AZ 85069}
\author{Monica Olvera de la
Cruz}\affiliation{Department of Materials Science and Engineering,
Northwestern University}
\date{\today}

\begin{abstract}
We consider a mixture of one neutral and two oppositely charged
types of molecules confined to a surface. Using analytical
techniques and molecular dynamics simulations, we construct the
phase diagram of the system and exhibit the coexistence between a
patterned solid phase and a charge-dilute phase. The patterns in
the solid phase arise from competition between short-range
immiscibility and long-range electrostatic attractions between the
charged species. The coexistence between phases leads to
observations of stable patterned domains immersed in a neutral
matrix background.
\end{abstract}

\maketitle

Mixtures of cationic and anionic amphiphilic molecules can form
thermodynamically stable structures such as micelles, membranes
and multilamellar systems. These self-assembled structures have
been studied as a function of the molar ratio of the oppositely
charged molecules, their concentration in solution, and the ionic
strength of the environment \cite{jokela,kaler2,brasher,Zemb2}. In
addition, the presence of other neutral components leads to a
large variety of structures and the possibility of local
organization on the surface of membranes and monolayers in
emulsions. Such structures are important in diverse applications,
such as the design bio-sensing devices \cite{biesalski}. Moreover,
they serve as model systems for the understanding of the
properties of cell membranes where structured domains are known to
be crucial to cell signaling processes \cite{groves}. In
multicomponent membranes periodic nanostructures can be immersed
in a homogenous background. This letter discusses the possibility
of this coexistence of periodic nanostructures with dilute phases
of immicible cationic and anionic molecules.  Besides charges on
biological membranes, the competition of short range immiscibility
and long range attractions leads to the formation of periodic
structures in a large variety of systems including lipid mixtures
\cite{seul,keller,mcconnell,andelman,heckl}, two dimensional
uniaxial ferromagnets \cite{garel}, reaction controled phase
segregating mixtures \cite{glotzer}, and two dimensional electron
gases in MOSFET's \cite{kivelson05}.

Phase separation phenomena at surfaces is well known
\cite{baumgart, veatch, zhang} and can be studied through simple
models of immiscibility. On the other hand, several recent studies
have shown that mixtures of immiscible oppositely charged
molecules can form regular periodic nanostructures (or
microphases) \cite{solis, ramos, zasadzinski,baksh}. To study the
convergence of these two phenomena we use analytic techniques as
well as off-lattice molecular dynamics simulations of a
coarse-grained model with two immiscible charged molecular
components. We assume that the membrane surface is in equilibrium
and that fluctuations perpendicular to the surface are negligible.
To begin with, we explore the phase behavior of this model
analytically when the domains are well segregated and periodically
ordered.  We briefly describe the effect of different charge
ratios of the molecules, but concentrate on the effects of the
presence of the third neutral component and determine the
conditions of coexistence between a microstructured solid and a
dilute gas of charges. Secondly, we describe the results of
molecular dynamics simulations for intermediate segregation
regimes. Simulation results exhibit the expected microscopic phase
behavior, but also indicate the limits of our theoretical
analysis.

\begin{figure}\begin{center}
\includegraphics[width=8.5cm]{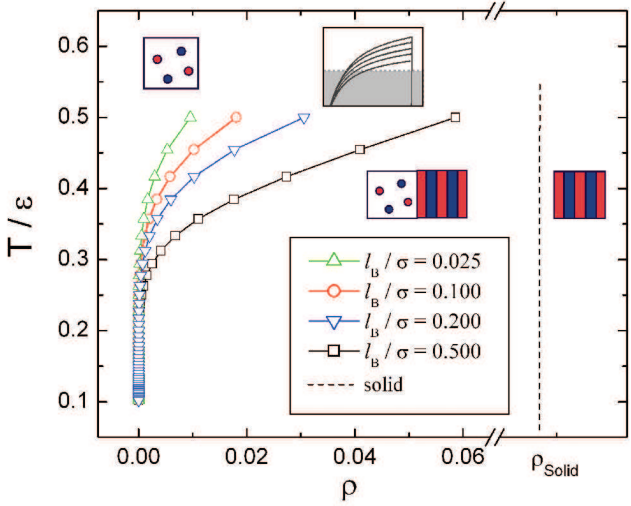}
\caption{\label{fig:figure1} Phase-diagram of the oppositely
charged immiscible mixture. Location of solid-gas coexistence
curves in $\rho-T$ plane is shown for several different values of
the Bjerrum length, $l_{B}$. Increasing the Bjerrum length
requires higher values of short range interactions $\epsilon$ for
phase coexistence. The curves shown for the left branch of the
coexistence region lie in the shaded region of the sketched full
diagram in the central inset. The solid phase has constant
density $\rho_{solid}$. Figure refers to a symmetrically charged
mixture that generates lamellar structures in the solid phase.  }
\end{center}
\end{figure}

We consider only two possible phases for the system at low
temperature. One phase consists of a dense, patterned solid
formed by the charged components. Its free energy is computed by
assuming the formation of regions of constant particle and charge
density. We ignore the fluctuations in the charge density and the
shape of the interface. The second phase is homogenous, has a low
density of charged particles, and can be treated as a two
dimensional charged gas. Its free energy can be calculated at low
temperatures using linear response theory by means of the Random
Phase Approximation (RPA). The resulting phase diagram is plotted
in Fig. \ref{fig:figure1}.

Consider first the calculation of the free energy of the solid
phase. The average absolute value of the charge density in this
phase is $\pm \psi$, (in electronic charges per unit area), and
the line tension between microdomains is $\gamma$. As shown in
previous work, the domains form a lamellar structure for near
symmetric charge density ratios, while for highly asymmetric cases
we have near-circular domains arranged in a hexagonal lattice. The
periodicity of both of these possible structures defines a
characteristic length $L$. Then, the free energy contribution per
area $A$, due to the microphase formation has the form
\begin{eqnarray}
\frac{F_{m}}{Ak_BT}= \gamma \frac{s_{1}}{L}+l_B \psi^{2} s_{2}L.
\end{eqnarray}
Here, the Bjerrum length is $l_{B} = \frac{e^2}{4\pi \varepsilon
k_{B}T}$, where $k_B$ is the Boltzmann constant and $T$ is the
temperature, and $\varepsilon$ is the dielectric permittivity of
the medium. The coefficients $s_{1}$ and $s_{2}$ are dimensionless
quantities that depend on specific shape of the microdomains.
$s_{1}$ is the ratio of the microdomain interface length within a
unit cell to the size of the cell. $s_{2}$ is the integral of the
dimensionless Coulombic potential over the whole space, averaged
over a unit cell. These values are calculated explicitly in
Ref.\cite{solis} for both the lamellar and hexagonal cases.
Minimizing the free energy with respect to the size $L$, we obtain
an optimal characteristic length $L_{min}=(s_1/s_2)^{1/2}L_{0}$,
with $L_{0}=(\gamma/(l_{B}\psi^2))^{1/2}$. Evaluation of the free
energy density at that point results on a value of
$2(s_{1}s_{2})^{1/2}f_{0}/L_{0}^2$, where $f_{0}=(\gamma l_{B}
\psi^2)^{1/2}$ \cite{solis,loverde}. The net free energy of the
solid phase also includes the cohesive energy that arises from
segregation of the charged molecules from the neutral component.
To be able to compare energies with the more dilute ionic gas, and
with results of the molecular dynamics simulations, it is
necessary to introduce a model of the individual molecules. In the
simulation, their Wan der Waals interactions are described by a
classic Lennard Jones potential
\begin{eqnarray}
U_{LJ} = 4\epsilon
\left(\left(\frac{\sigma}{r}\right)^{12}-\left(\frac{\sigma}{r}\right)^6\right)&&
r<r_{c}
\end{eqnarray}
where the potential is cut at a radius $r_{c}=2.5\sigma$ for
oppositely charged molecules and $r_{c} = 2 ^{\frac{1}{6}}$ for
similarly charged molecules, where $\sigma$ is an effective
molecular radius. This selection of cut-offs produces a net
immiscibility between species of magnitude $\epsilon$ per contact
between oppositely charged particles. In our description of the
solid phase, at low temperatures, we simply assume a constant
density given by the hcp packing of spherical molecules of radius
$\sigma$. For this close packing arrangement we have the constant
density $\rho_{solid}=\frac{1}{\sqrt{3}\sigma^{2}}$. The effective
cohesive energy per unit area can then be written in terms of
$\epsilon$, with 6 favorable contacts with neighbors. Using a
similar approximation, the effective line tension is
$\sqrt{6}\epsilon/\sigma$. Inclusion of the cohesive energy, leads
the final result for the solid phase as:
\begin{equation}
\frac{F_{s}(\rho_{solid})}{A}=-3\epsilon\rho_{solid}+2(s_{1}s_{2})^{1/2}f_{0}/L_{0}^2.
\end{equation}

Next, consider the free energy of the gas phase.  RPA
\cite{gonzalez2} is used to calculate the free energy as a
function of the relative strength of the short range attraction
and the electrostatic interactions. The method involves an
expansion of the free energy of the system in terms of density
fluctuations, neglecting all terms of larger than second order.
For a general system of interacting $i,j$ charged monomers, the
partition function is written as
\begin{eqnarray}
Z = Z_{o}\frac{V^{N}}{N!}\int exp\left(-\sum_{k\neq
0}\sum_{ij}\frac{(\textbf{U}_{k}^{ij}+\rho_{i}^{-1}\delta_{ij})\rho_{k}^{i}\rho_{-k}^{j}}{2V}\right)\\\times\prod_{k>0}\prod_{i}\frac{d\rho_{k}^{i}}{\pi
V \rho_{i}}\nonumber.
\end{eqnarray}
Here, $\rho_{i}$ represents the density of the $i$th component
and $\rho_{k}$ represents the Fourier transform of the component
densities. $Z_{o}$ includes the $k = 0$ and self energy terms.
$\textbf{U}_{k}^{ij}$ is the sum of the interaction energies,
consisting of the short range interactions due to the excluded
volume and hydrophobic interaction, as well as the long range
contributions due to the electrostatic energy. The electrostatic
contribution to the internal energy matrix uses the two
dimensional Fourier transform of the screened Coulomb
interaction, $E_{el}^{ij}=2\pi
z_{i}z_{j}l_{B}(k^{2}+\kappa^{2})^{-1/2}$. The free energy is
then
\begin{eqnarray}F=\sum_{ij}\frac{N_{i}(N_{j}-\delta_{ij}))}{2V}U_{o}^{ij}+\sum_{i}N_{i}\ln{\frac{\rho_{i}}{e}}\\+\sum_{k>o}\left[\ln{\frac{det|U_{k}^{ij}+\rho_{i}^{-1}\delta_{ij}|}{det|\rho_{i}^{-1}\delta_{ij}|}}-\sum_{i}\rho_{i}U_{k}^{ii}\right]\nonumber.
\end{eqnarray}
The first sum (the $k = 0$ terms) vanishes due to charge
neutrality.  The second term is a standard entropic term. The
third term is the electrostatic contribution due to density
fluctuations, which is calculated by integrating over the
possible values of $k$ from $0$ to $2\pi/a$, where $a$ is the
molecular size. We neglect terms arising from excluded volume
interactions, as the calculation is dominated by the
electrostatic terms. We only consider the limit of no screening
due to ions in the surrounding solution (matching the conditions
of the calculation for the solid phase and for the molecular
dynamics simulations). The electrostatic contribution is found to
be,
\begin{eqnarray}
\frac{F_{el}}{k_{B}T}=\frac{1}{4\pi}\left[\frac{2\pi^2\ln(1+\frac{ak_{in}^2}{2\pi})}{a^2}+\frac{\pi
k_{in}^2}{a}\right.\\\left.-\frac{1}{2}k_{in}^4\ln(1+\frac{2\pi}{ak_{in}^2})\right]\nonumber\end{eqnarray}
where $k_{in}^{2}=2\pi
l_{B}(\rho_{+}z_{+}^{2}+\rho_{+}z_{+}^{2})$. The above equations
simplify when considering the charge neutrality constraint,
$z_{+}\rho_{+}=z_{-}\rho_{-}.$    The total free energy per unit
area for the gas phase, in terms of $\rho$ where $\rho =
\rho_{+}+\rho_{-}$, is then
\begin{eqnarray}\frac{F_{gas}}{Ak_{B}T}=\frac{\rho}{\alpha}\ln\left[\frac{\rho}{\alpha e}\right]+\frac{\rho}{\beta}\ln\left[\frac{\rho}{\beta
e}\right]\\+\frac{\upsilon_{11}}{2}\left(\frac{\rho}{\alpha}\right)^{2}+\frac{\upsilon_{22}}{2}\left(\frac{\rho}{\beta}\right)^{2}+\upsilon_{12}\frac{\rho^{2}}{\alpha
\beta}+\frac{F_{el}}{k_{B}T}\nonumber
\end{eqnarray} where $\alpha=1-\frac{z_+}{z_-}$ and $\beta=1-\frac{z_-}{z_+}$. The virial terms are
$\upsilon_{ij}= -\int e^{-U_{ij}/k_{B}T }-1$ where $U_{ij}$ is a
hard core potential from $0<r<\sigma$ and a classic 6-12 Lennard
Jones potential from $\sigma<r<2.5\sigma$.

To determine the phase coexistence of the solid, patterned phase
and the dilute gas, we use the common tangent rule. Since the
density of the solid phase is assumed fixed, the equation
\begin{equation}
F_{S}\left(\frac{\rho_{solid}}{A}\right)
=\frac{F_{gas}}{A}+(\rho_{solid}-\rho)\frac{\partial
F_{gas}}{\partial \rho}
\end{equation}
is solved for the gas phase density $\rho$. We plot the two phase
coexistence line calculated in this way, for low temperatures, in
Fig. 1. The figure refers to the case of equal charge density
$z_{+}/z_{-} = 1$. The phase diagram shape depends of course on
the relative strengths of the Coulomb interaction (through the
charge density) and on cohesive/immiscibility parameter
$\epsilon$. With increasing strength of the electrostatics, the
transition points occurs at lower temperatures,  while increasing
values of the short range interaction increase those
temperatures.  At higher values of the density, nonlinear
corrections to RPA including short range correlations and ion
association \cite{gonzalez3, ermoshkinmacro}. Fig. 1 sketches an
extrapolation of the results to higher temperatures when the
solid phase retains its near constant density.

To explore higher temperatures and the effects of fluctuations,
we carried out a set of molecular dynamics simulations. The model
systems used are composed, in the symmetric case, of a mix of
$N_{+}=1000$ positively charged and $N_{-}=1000$ negatively
charged units in a simulation box of size $D^{3}$, with $D=66
\sigma$. In the asymmetric case we used $N_{+}=900$ and
$N_{3-}=300$ units with charge (-3). The molecules are confined
to a plane perpendicular to the Z axis, with periodic boundary
conditions in the X and Y directions. The molecular dynamics
simulations, with constant N,V,T were performed using the {\it
Espresso} simulation code of the MPIP-Mainz group of Polymer
Theory and Simulation \cite{limbach}. We explored regions of the
phase diagram at surface densities of $\rho = \frac{(N_{+} +
N_{-})\pi \sigma^{2} }{4 D^{2}} = 0.36$. The potential between
charges is a Coulomb potential, $U_{C} =
\frac{l_{B}Tq_{1}q_{2}}{r}$.  Further parameters are as described
in detail in previously published work \cite{loverde}.
\begin{figure}
\begin{center}
\includegraphics[width=8.6cm]{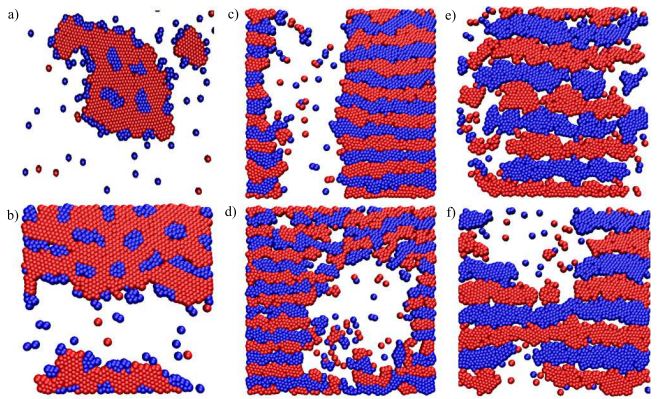}
\caption{\label{fig2} Simulation snapshots for systems with charge
ratios, a-b) $z_{+}/z_{-} = 1/3$ and c-f) $z_{+}/z_{-} = 1$.
Frames a-b illustrate hexagonal order for different densities
$\rho_{a} = 0.10,\rho_{b} = 0.36$. Frames c-d show the effect of
increased immiscibility $\epsilon_c= 3.4k_BT$,
$\epsilon_d=3.4k_BT$, for a fixed Bjerrum length, $l_{B}/\sigma =
0.5$. Frames e-f show the transition from a homogeneous microphase
to a phase segregated state for  $\epsilon_e = 2.0 k_BT $,
$\epsilon_f= 2.8 k_BT$.}
\end{center}
\end{figure}
Late-time snapshots  of the simulations are shown in Fig. 2. To
compare the behavior observed in the simulations to our
theoretical results, we point out that the simplest extrapolation
of the results from the strong segregation limit to higher
temperatures, consist on identifying the characteristic length
scale of system (the size of microdomains), up to a constant, with
the form $L=(\gamma/(l_{B}\psi^2))^{1/2}
\sim(\epsilon/(l_{B}\psi^{3/2}))^{1/2}$. To use this expression
for extrapolation, we abandon the assumption of a solid phase
density fixed by the molecular radius, but take $\psi$ as the
density determined by the simulation at finite temperatures. All
our simulations exhibit the same expected basic behavior. At small
values of $\epsilon$ or low Bjerrum lengths (high temperature),
positive and negative regions develop on the surface and as the
temperature decreases the domains increase in size. In all the
simulations, the individual molecular components exhibit a
hexagonally close-packed structure (as was assumed in our
analytical approach), with density fluctuations dependent on the
temperature. For asymmetric charge ratios, $\frac{z_{+}}{z_{-}} =
1/3$, we show in Fig. 2a  the formation of a hexagonally patterned
'island' at $\rho = 0.10$. For larger densities, as in Fig. 2b the
solid phase occupies a larger fraction of the space, but exhibits
more clearly the ordering.

Next, as shown in Fig. 2c-d, for symmetric charge ratios,
$\frac{z_{+}}{z_{-}} = 1$ the preferred microstructure is
lamellar. At the values of the parameters used, we observe as well
the phase separation between solid patterned and neutral regions.
The temperature decrease from snapshot {\it c} to {\it d}, clearly
modifies the fluctuations of both types of interfaces: between
charged regions, and between the solid and neutral phases. At
lower temperatures (Fig. 2d) the interfaces are much sharper and
exhibit lower shape fluctuations. In these simulations, a note and
an interesting feature of the solid-gas interface: the orientation
of the lamella is perpendicular to the interface. While the
charged domains have symmetric interactions with the neutral
region, the alignment must be a result of minimization of the
local electrostatic energy. To some extent, this feature also
appears in the asymmetric case, Fig. 2a-b, where both charged
domains appear the at the interface wit the neutral domain.

For symmetric charge ratios, but for weaker electrostatic
interactions, Fig. 2e-f show the transition from a solid to a
solid-gas coexistence phase.  At low values of the cohesive energy
$\epsilon$ the system shows the lamellar patterning but has large
voids between the charged domains; the neutral component is
attracted to the interfaces where it can reduce the effective line
tension between domains. On further increase in the cohesive
parameter, the coexistence region is reached, and the neutral
regions segregate to form their own phase, as shown in Fig. 2f.
The lower values of the Bjerrum length in these cases produce
larger lamellar sizes, compared with those of Fig. 2c-d.

\begin{figure}[t]
\includegraphics[width=8.5cm]{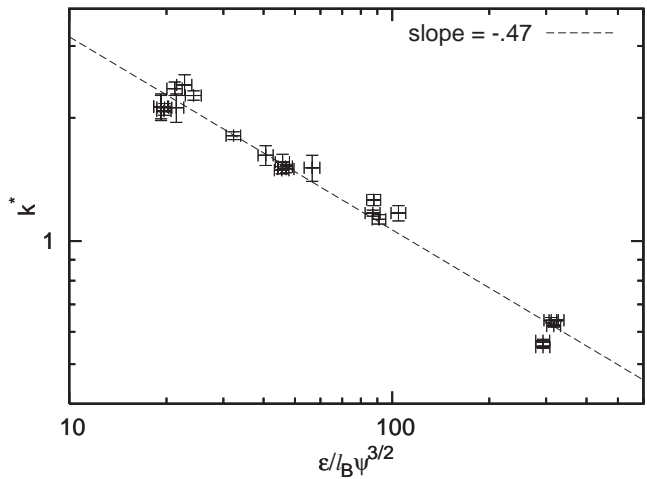}
\caption{\label{fig3} The location of the peak $k^{*}$ in the
structure factor $S(\vec{k})$ as a function of
$\epsilon/(l_B\psi^{3/2})$. The linear fit shows agreement with
the scaling predicted by strong segregation theory of power
$-0.5$. }
\end{figure}
To study the behavior of the domain size (lamellar width) in the
symmetric case, we have determined the structure factor
$S(\vec{k}) = \left<\rho_{k}\rho_{-k}\right>$, the Fourier
transform of the correlation function of the charged components,
for the late stages of a number of simulations with different
values of the interaction parameters. $S(\vec{k})$ displays a peak
at values $k^{*}$ corresponding to the inverse lamellar spacing in
the direction perpendicular to the lamellas, and thus must scale
as $k^{*}\sim(\epsilon/(l_{B}\psi^{3/2}))^{-1/2}$. Indeed, in our
simulations, we find that the position of the structure factor
peak can be fit through a line of slope $-.47 +/- .02$ when
plotted against the group $\epsilon/(l_{B}\psi^{3/2}$), as shown
in Fig. 3.

By use of analytic techniques, combining RPA and strong
segregation theory, as well as off-lattice molecular dynamics
simulations we have demonstrated the clear possibility of
coexistence of structured regions, made of charged components, in
a matrix rich in neutral components. The competition between
immiscibility tendencies between molecules, against the attractive
interactions between oppositely charged species, provides a method
to generate well-controlled, self-assembled surface patterns.
Furthermore, suitable extensions of the results from the strong
segregation limit to higher temperatures provide a clear way to
quantitatively address the properties of these materials in
practical regimes.

This work was supported by the IGERT-NSF Fellowship awarded to S.
Loverde, and by grant numbers DMR-0414446 and EE-0503943.  We
acknowledge helpful discussions Yury Velichko, Graziano Vernizzi,
Christian Holm, including Axel Arnold and Bernward Mann.  S.
Loverde thanks the hospitality of the group of Christian Holm
where this work was initiated.

\end{document}